\documentclass[aps,prl, twocolumn, superscriptaddress]{revtex4-1}
\usepackage{graphicx}
\usepackage{amsmath}
\usepackage{amsfonts}
\usepackage{amsthm}
\usepackage{amssymb}
\usepackage{amsbsy}
\usepackage{wasysym}
\usepackage{bm}
\usepackage{mathrsfs}
\usepackage{color}
\usepackage{times}
\usepackage{subfig}
\newcommand{\beq}{\begin{equation}}
\newcommand{\eeq}{\end{equation}}
\newcommand{\bea}{\begin{eqnarray}}
\newcommand{\eea}{\end{eqnarray}}

\def\zhat{\hat{\mathbf{z}}}

\def\ket#1{{\left|#1\right\rangle}}
\def\bra#1{{\left\langle #1 \right|}}

\def\br{\mathbf{r}}
\def\bk{\mathbf{k}}
\def\bq{\mathbf{q}}
\def\bx{\mathbf{x}}

\def\be{\begin{eqnarray}}
\def\ee{\end{eqnarray}}

\begin{document}

\title{Dipolar Bogolons: From Superfluids to Pfaffians}
\author {S. A. Parameswaran}
\email{sidp@berkeley.edu}
\affiliation{Department of Physics, University of California, Berkeley, CA 94720, USA}
\author {S. A. Kivelson}
%\email{kivelson@stanford.edu}
\affiliation{Department of Physics, Stanford University, Stanford, CA 94305, USA}
\author {R. Shankar}
\affiliation{Department of Physics,  Yale University, New Haven, CT 06520, USA}
\author {S. L. Sondhi}
%\email{sondhi@princeton.edu}
\affiliation{Department of Physics, Princeton University, Princeton, NJ 08544, USA}
\author {B. Z. Spivak}
%\email{spivak@u.washington.edu}
\affiliation{Department of Physics, University of Washington, Seattle, WA 98195, USA}
\date{\today}

\date{\today}
\begin{abstract}
We study the structure of 
 Bogoliubov quasiparticles,  `bogolons,'  the  fermionic  excitations of paired superfluids that arise from fermion (BCS) pairing,  including neutral superfluids,
superconductors, and paired quantum Hall states.
The na\"{i}ve construction of a stationary quasiparticle in which the deformation of the pair field is neglected leads to a contradiction: it carries a net
electrical current {\it even though it does not move.} However, treating the pair field self-consistently resolves this problem:  In a neutral superfluid,
a dipolar current pattern is associated  with the  quasiparticle for which the
total current vanishes.  When Maxwell electrodynamics is included, as appropriate to a superconductor, this pattern is confined over a penetration depth. For paired quantum Hall states of composite fermions, the Maxwell term is replaced by a Chern-Simons term, which leads to a dipolar {\it charge}  distribution and consequently to a dipolar current pattern. \end{abstract}
\maketitle

 {\it Introduction.---} Paired superfluids are among the most ubiquitous of the many ordered phases
of interacting fermions in two and three dimensions. In condensed matter settings they include both neutral superfluids such as $^3$He, charged superconductors and now also paired quantum Hall (QH) liquids such as the Moore-Read or Pfaffian state that is believe to underlie the quantized Hall plateau at filling factor $\nu=5/2$. In all cases paired superfluids exhibit two distinct excitations that dominate much of their
physics: vortices and Bogoliubov quasiparticles or `bogolons'. The former are a generic consequence
of superfluidity but the latter are a particular
signature of pairing--- 
 they involve breaking apart a Cooper pair into its fermionic constituents.

The structure of vortices is well understood:
They are the topological solitons of a complex scalar order parameter in the Landau-Ginzburg description of a superfluid.  In a superconductor, additional  coupling to a Maxwell gauge field results in
an associated  quantum of flux, while for a quantum Hall liquid, coupling to a Chern-Simons gauge field associates a quantized charge with each vortex.
 The structure of bogolons is  less
 well understood as they are, by comparison, much more quantum mechanical particles. We will address
that gap by providing a theoretical analysis of their structure for 
all three examples alluded to above. For superfluids and superconductors
we will be able to recover the heuristic description advanced by  Kivelson and Rokhsar
\cite{Kivelson:1990p1}.
For paired quantum Hall liquids our results
are new and add to a recent burst of interest in the properties of bogolons \cite{Bonderson:2010p340,Moller:2010p1370}, including work
by four of the present authors \cite{Parameswaran:2011p1}.

 In the weak pairing (BCS) limit
 the   momentum (or Bloch) eigenstates of the bogolon exhibit the well known dispersion relation sketched in Figure 1a, with a characteristic minimum at the underlying Fermi surface.
In terms of these, one can make a localized wave packet state with a spatial extent large compared to the coherence length, $\xi$,  and a well defined momentum. 
Unlike a wavepacket in the normal state, this bogolon wavepacket has a
group velocity which is different than the Fermi velocity $v_{F}$, and which vanishes on the Fermi surface.
It has spin $1/2$, but its (average) charge is smaller than the charge of electron $e$.
Both quantities vanish as the momentum of quasiparticles $p$ approach the Fermi momentum $p_{F}$.
 On the other hand, since the wavepacket has a net momentum, it
 carries a net current \cite{Nayak:2001p1} equal to $ev_{F}$. This indicates that our
construction of a localized bogolon is fundamentally inadequate. The problem becomes especially clear in the limit $p=p_{F}$, where the group velocity of the wavepacket is zero. In this case the current density is finite inside the wavepacket and zero outside of it. The resolution of this puzzle
will lead us to a bogolon structure that involves an algebraically falling, dipolar, return current flow via
the condensate for neutral superfluids, a version of this screened on the scale of the London
length for superconductors, and a version exhibiting a charge dipole as well as a
locally dipolar backflow for two dimensional quantum Hall fluids. Altogether,
bogolons are fairly complicated objects!

\begin{figure}
\subfloat[]{%
\begin{minipage}[c][1\width]{1.7in}%
\includegraphics[clip,width=1\textwidth]{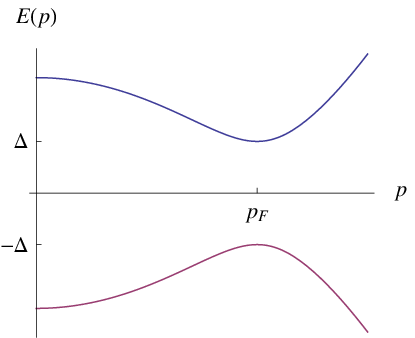}%
\end{minipage}}\subfloat[]{\centering{}%
\begin{minipage}[c][1\width]{1.7in}%
\begin{center}
\includegraphics[clip,width=1\textwidth]{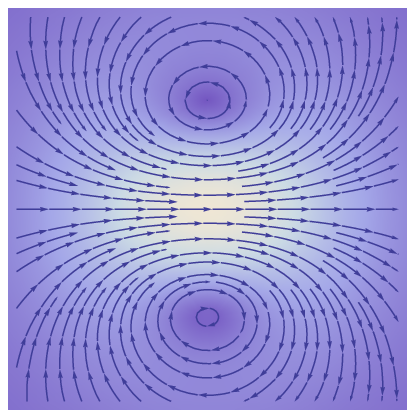}
\par\end{center}%
\end{minipage}}
\caption{(a) Quasiparticle dispersion. (b) Current flow around neutral superfluid bogolon (window size $\sim 2\lambda$.)}
\end{figure}

 {\it Bogolon wavepacket.---}
We begin with the mean-field BCS Hamiltonian for a neutral
fully gapped paired superfluid which also
serves to fix notation,\be
H_{\text{\tiny BCS}} =  \sum_{\substack{\bk\\ s = \uparrow,\downarrow}} \xi_{\bk} c^\dagger_{\bk s} c_{\bk s} + \sum_{\bk}\left [ \Delta_\bk c^\dagger_{\bk\uparrow} c^\dagger_{-\bk\downarrow}+{\rm h. c.}\right]
\ee
where $\xi_\bk = \frac{k^2}{2m}-\mu$,  $\Delta_\bk$ is the gap function
and we work in units where $\hbar=e/c=1$.
 It is a simple matter to diagonalize, $H_\text{\tiny BCS} = \sum_{\bk,s} E_{\bk} \gamma^\dagger_{\bk,s}\gamma_{\bk,s}$, with $E_{\bk} = \sqrt{{\xi}_\bk^2 + |\Delta_\bk|^2} $  by means of a Bogoliubov transformation,
\be\label{eq:BogTrans}
\gamma_{\bk,\uparrow} = u_\bk c_{\bk,\uparrow} - v_{\bk} c^\dagger_{-\bk,\downarrow},& &
\gamma_{\bk,\downarrow} = v_\bk c^\dagger_{\bk,\uparrow} + u_{\bk} c_{-\bk,\downarrow}, \nonumber\\
|u_\bk|^2 = \frac{1}{2} \left(1 + \frac{\xi_\bk}{E_\bk}\right), & & |v_\bk|^2 = \frac{1}{2} \left(1 - \frac{\xi_\bk}{E_\bk}\right).\ee
The ($T=0$) BCS ground state is then the state annihilated by all the $\gamma_{\bk s}$, $\ket{\Omega}  = \prod_{\bk \ge 0} \left(u_\bk + v_\bk c^\dagger_{\bk\uparrow}c^\dagger_{-\bk\downarrow} \right)\ket{0}$. Single-quasiparticle states with momentum $\bk$ are  given by $\ket{\bk s} = \gamma^\dagger_{\bk s} \ket{\Omega}$ and it is readily verified that their energy $E_\bk$
is minimal
at $|\bk| = k_F$.
We will work in $d=2$ as that naturally includes the case of the paired QH state, but the results are readily generalized to $d=3$.

A quasiparticle wavepacket  with average momentum $\hbar\bk_0 =  \hbar k_F \hat{\bk}_0$, spin $s$ and spatial extent
$\sim \lambda$ is obtained by superposing the states $\ket{\bk s}$ with momenta near $\bk_0$
\be\label{eq:wpstate}
\ket{\Psi^\lambda_{\bk_0,s}} = \left(\frac{\lambda}{\sqrt{\pi}}\right)^{\frac{d}{2}}\int{d^dk}\, e^{-\frac12\lambda^2(\bk - \bk_0)^2} \ket{\bk s}.
\ee
In order that the energy uncertainty of the wavepacket be smaller than its average energy, we need
to choose $\lambda 
\gg \xi={v_F \over \Delta_0}$ as can be deduced from the low lying dispersion relation
$E(\bk) \approx \Delta_0 +\frac {[v_F\hat{\bk}_0\cdot(\bk-\bk_0)]^2} {2\Delta_0}$ where $\Delta_0\equiv |\Delta_{\bk_0}|$ and $v_F = k_F/m$ is the Fermi velocity. 

Our primary concern is  the structure of quasiparticle wavepackets centered at momenta close to $p_{F}$, so that their group velocity is much smaller than $v_{F}$. Clearly the packet  has vanishing group velocity at $p=p_F$. However,  a tedious but straightforward computation of the expectation value of the quasiparticle current operator $\mathbf{j}^{ \text{qp}}_\bq = \sum_{\bk,s} \frac{\bk}{m} c^\dagger_{\bk +\frac{\bq}{2} s} c_{\bk -\frac{\bq}{2} s}$ in the state yields \footnote{To derive this result we rewrite $c^\dagger, c$ in terms of $\gamma^\dagger, \gamma$ and expand $u_\bk, v_\bk$ around $k_F$ to linear order.}
\be
\langle\mathbf{j}^\text{qp}_\bq\rangle_\Psi =v_F \hat{\bk}_0 \ e^{- \frac{\lambda^2 \bq^2}{4} }
\label{jF}
\ee
We are thus presented with a contradiction: a {\it stationary} quasiparticle wavepacket is associated with a current that has nonzero divergence---violating the continuity equation.

A first step in resolving this puzzle is to observe that we have taken a slippery step in
passing from momentum space to real space. In real space, the wavepacket state (\ref{eq:wpstate}) is
now inhomogeneous and hence a homogenous ``pair potential''  $\Delta$ no longer yields a
self-consistent mean field theory of the wavepacket \cite{Bishop:1988p1}. It is possible to prove that any state that satisfies the self-consistency conditions respects the equation of continuity. Recomputing the pair potential in the wavepacket state and then iterating the construction of the wavepacket
and the computation of the pair potential should yield a state that does obey current
conservation \cite{Parameswaran:2012unpub}. In the supplementary material,
 we show that the first iteration of this process  produces a change in the pair potential that already partially cancels
the quasiparticle current.

However implementing this approach requires detailed numerical work. Instead, we
construct an effective action which correctly treats the low-energy, long-wave-length physics in the weak coupling limit, $\Delta_0 \ll E_F=k_F^2/2m$.
 While portions of this work may be reconstructed from existing literature, in particular the `conserving approximations' \cite{Anderson:1958p1, Pines:1958p1, Rickayzen:1959p1,Nambu:1960p1, BaymKadanoff} to superconducting response, to our knowledge an explicit quantitative treatment of a bogolon wavepacket has not been previously presented.

 {\it Neutral superfluids.---} As we are interested in a wavepacket constructed from momenta very
close to the (parent) Fermi surface, it is sufficient that we work with the effective dynamics
for this set of degrees of freedom. Formally, we begin with a Hubbard-Stratonovich (HS)
decoupling of an attractive four-fermion interaction in the particle-particle channel, and
integrate out fermions above a cutoff thus generating an effective action for the HS field,
$\Delta(\br,t)= \Delta_0 e^{i\theta(\br,t)}$.
 As we are in the broken-symmetry phase,
 fluctuations of the amplitude can be neglected.  The
 result is an effective theory
 of dynamical fermions coupled to a dynamical phase field $\theta(\br,t)$ \cite{AltlandSimons}.

To be explicit, we consider the case of s-wave pairing, where the most important terms in this (well known) theory
are represented by the action
$S = \int dt d^2r \, (\mathcal{L}_\psi +\mathcal{L}_{p} +\mathcal{L}_{\theta})$,
with\be
\mathcal{L}_\psi = \sum_s \psi_s^\dagger(\br,t)\left[i \partial_t - \mu -\frac{\nabla^2}{2m}\right]\psi_s(\br,t)\ ,
\ee
\be
\mathcal{L}_p &=& -\Delta_0e^{i \theta (\br,t)}\psi^\dagger_\uparrow(\br,t) \psi^\dagger_\downarrow(\br,t) + \text{h.c.}\ ,
\ee
\be
\mathcal{L}_\theta  = -\frac {\chi_0} 2 (\partial_t \theta)^2 + \frac{n_s}{2m} (\nabla\theta)^2\ ,
\ee
 where $\chi_0$  is the  static compressibility (equal to the density of states at the Fermi surface), and $n_s$ is the superfluid density. At $T=0$, $n_s= \rho$, the total electronic density, for the Galilean invariant systems considered here.

Note that the conserved charge is no longer carried solely by the fermions, but also by the superfluid component via
twists in the order parameter. A straightforward application of the Noether  procedure allows us to write, for the density  and current
\be \label{eq:conservedqties}
\rho = \rho^\text{qp} - \chi_0 \partial_t \theta, \,\,\, \mathbf{j} = \mathbf{j}^{ \text{qp}} +\frac{n_s}{2m} \nabla\theta \ee
where $\rho^\text{qp} =  \sum_s\psi^\dagger_s\psi_s$ and $\mathbf{j}^\text{qp} = \sum_s \frac{1}{m}\text{Im}[\psi^\dagger_s \nabla \psi_s]$.
From $S$ we then obtain the equations of motion
\be
\partial_t \rho^\text{qp} &=&  -\nabla\cdot\mathbf{j}^{ \text{qp}} + \mathcal{B}_{p} \label{eq:eom1}\\
\chi_0\partial_t^2 \theta &=&\frac{n_s}{2m}\nabla^2 \theta  + \mathcal{B}_{p} \label{eq:eom2}
\ee
where
$\mathcal{B}_{p} \equiv 2i\Delta_0 \left(e^{i\theta}\psi^\dagger_\uparrow \psi^\dagger_\downarrow - e^{-i\theta} \psi_\downarrow\psi_\uparrow \right)$
is  the term that couples the quasiparticles and the superfluid. From (\ref{eq:conservedqties}), (\ref{eq:eom1}) and (\ref{eq:eom2}),  it is evident that $\partial_t \rho + \nabla \cdot\mathbf{j} =0$, i.e. the properly defined density and current obey the continuity equation; it is equally clear that the quasiparticle density is {\it not} independently conserved.

Let us now specialize to the treatment of
a stationary bogolon wavepacket in the approximation where we
ignore the quantum fluctuations of $\theta$.
 This implies that the LHS of Eqns (\ref{eq:eom1}) and (\ref{eq:eom2})
 vanish so that
\be\langle \nabla\cdot\mathbf{j}^{ \text{qp}} \rangle = \langle \mathcal{B}_{p} \rangle 
=-\frac {n_s}{2m} \langle \nabla^2\theta\rangle.\ee
Thus, in the wave packet state for which $\langle\mathbf{j}^\text{qp}_\bq\rangle_\Psi$ is given by Eq. \ref{jF}, the resulting phase texture is
\be
\left\langle
\theta_\bq \right\rangle_\Psi=\frac{i (\bq\cdot\bk_0)
}{q^2} \left(\frac {2k_F} {n_s}\right) e^{-\frac{\lambda^2\bq^2}{4}},
\ee
which permits us to write for the total current
\be\label{eq:kspacecurrent}
\langle\mathbf{j}_\bq\rangle_\Psi =  v_F\left[\frac{q^2\hat{\bk}_0 -  (\bq\cdot\bk_0)\bq} {q^2}\right] e^{-\frac{\lambda^2 \bq^2}{4}}
\ee
 Eq. (\ref{eq:kspacecurrent}) corresponds to a real space current  $
\langle\mathbf{j}(\br)\rangle_\Psi = \zhat\times\nabla\varphi_\lambda(\br)$, where
$\varphi_\lambda(\br) \equiv 2\pi v_F \frac 
{(\hat{\bk}_0\times \br)\cdot\zhat}{r^2}\left(1 - e^{-r^2/\lambda^2}\right).$

The flow pattern is solenoidal (clearly $\nabla\cdot\langle\mathbf{j}\rangle_\Psi = 0$), and decays  as $r^{-2}$ far from the center of the wavepacket.
Corrections to this expression at short distances are non-universal, and are beyond the reach of the field-theory approach. Finally, we note that at {\it finite} quasiparticle concentration $\overline{\rho^\text{qp}}$, the long-range nature of the distribution of current density  in a quasiparticle wave packet leads to the conventional expression ${\bf j}^\text{qp}=e{\bf v}_{F}{\rho^\text{qp}}$ for the quasiparticle contribution to the current density,
 in agreement with the Boltzmann approach \cite{RussianReview} applicable in this limit.

 {\it Superconductors.---} We now
turn to  the case
of a charged superfluid
which is minimally coupled to a fluctuating $U(1)$ Maxwell gauge field $A_\mu$ -- i.e., the superconductor with dynamical electromagnetism.
The effective action is obtained from that of the neutral superfluid by converting the derivatives to  covariant derivatives: $\partial_\mu  \rightarrow D_\mu =  \partial_\mu - iA_\mu$,  
where the dynamics of $A_\mu$ are described by $\mathcal{L}_\text{Maxwell} =  \frac{1}{4} F_{\mu\nu} F^{\mu\nu}$ in which $F_{\mu\nu} = \partial_\mu A_\nu - \partial_\nu A_\mu$ is the Maxwell field strength. From $S +S_{\text{Maxwell}}$, we find the equations of motion for the quasiparticle and superfluid currents
\be\label{eq:eom_sc}
\rho = \rho^\text{qp} - \chi_0 (\partial_t \theta&-&2A_0), \,\,\, \mathbf{j} = \mathbf{j}^{ \text{qp}} +\frac{n_s}{2m} (\nabla\theta-2\mathbf{A}) \nonumber \\
\partial_t \rho^\text{qp} &=&  -\nabla\cdot\mathbf{j}^{ \text{qp}} + \mathcal{B}_{p} \nonumber\\ 
\chi_0\partial_t (\partial_t \theta - 2A_0)  &=& \frac{n_s}{2m} \nabla\cdot(\nabla\theta -2\mathbf{A}) +\mathcal{B}_{p} \ee
supplemented by Maxwell's equations
\be\label{eq:Maxwell}
\nabla \cdot \mathbf{E} = 4\pi (\rho -\bar\rho) &,&\,\,\,\,\,\, \nabla\cdot\mathbf{B} =0,  \nonumber\\
\nabla \times \mathbf{B} =  {4\pi}\mathbf{j}  + \partial_t \mathbf{E}&,&\,\,\,\,\,\, \nabla\times \mathbf{E} = -\partial_t\mathbf{B}.
\ee
In (\ref{eq:eom_sc}) and (\ref{eq:Maxwell}) the quasiparticle current and density take their gauge-invariant forms, $\rho^\text{qp} =  \sum_s\psi^\dagger_s\psi_s$ and $\mathbf{j}^\text{qp} = \sum_s \frac{1}{m}\text{Im}[\psi^\dagger_s \mathbf{D} \psi_s]$, and $\mathbf{E} = -\partial_t \mathbf{A} - \nabla A_0$ and $\mathbf{B} = \nabla\times \mathbf{A}$ are the electric and magnetic fields in the quasiparticle state; in writing the Poisson equation we have assumed the existence of a neutralizing positive background $\bar\rho = \langle \sum_s \psi^\dagger_s\psi_s\rangle_\Omega$ in the BCS ground state.

The first comment to be made here is that now even
extended bogolon states of definite momentum 
do not carry current. This basically reflects the Meissner effect.  Specifically,  the uniform quasiparticle contribution to the current  is
exactly cancelled by a superfluid backflow, which in unitary gauge $\theta=0$, corresponds to $\frac{n_s}{m}\mathbf{A} = \langle\mathbf{j}^{\text{qp}}\rangle\propto v_F \hat{\bk}_0$ \footnote{Here we have in mind bulk bogolon states. The momentum space description breaks down within a London length of any boundaries.}. The correct bogolon state carries no current; they are neutral particles.

Still in unitary gauge, let us turn to the construction of the wavepacket. For static wavepackets 
 we find that the 
 third equation of (\ref{eq:eom_sc}) yields $\frac{n_s}{m}\nabla\cdot\mathbf{A} =\langle \mathcal{B}_p\rangle
= \langle \nabla\cdot \mathbf{j}^\text{qp}\rangle$, so that as before the total current $\mathbf{j} =  \mathbf{j}^\text{qp} - \frac{n_s}{m}\mathbf{A}$ is conserved. Using this, we rewrite the third Maxwell equation as
\be
\left[ -\nabla^2 + \lambda_L^{-2}\right] \mathbf{A} &=& 4\pi\left\langle \mathbf{j}^\text{qp} - {\lambda_L^{2}} \nabla(\nabla\cdot\mathbf{j}^\text{qp})\right\rangle \label{eq:MaxJ} 
\ee
where we have defined the penetration depth via $\lambda_L^{-2} =  \frac{4\pi n_s}{m}$ and the  expectation value is taken in the naive wave packet state with $A_0 = \mathbf{A} =0$. 
Using either the expectation value of $\langle\mathcal{B}_p\rangle$ computed in the superfluid case or the form of $\mathbf{j}^\text{qp}$, we may solve (\ref{eq:MaxJ}) by Fourier analysis: 
\be\label{eq:Maxkspacecurrent}
\langle \mathbf{j}_\bq \rangle_\Psi = v_F
\left[\frac{q^2\hat{\bk}_0 -  (\bq\cdot\hat{\bk}_0)\bq} 
{q^2 + \lambda_L^{-2}}\right] e^{-\frac{\lambda^2 \bq^2}{4}} ,
\ee
which coincides with (\ref{eq:kspacecurrent}) in the limit in which the coupling to electromagnetism vanishes (when $\lambda_L \rightarrow\infty$).
It is easy to see that the power-law asymptotics of the superfluid case are replaced by exponential behavior at long distances, $\mathbf{j}(\br) \sim e^{-r/\lambda_L}$ for $r\gg \lambda$ and $\lambda_L$.
 This reflects the fact that superconductors screen magnetic fields and thus the current pattern is confined to within a penetration depth of the center of the bogolon \footnote{We remark that the otherwise excellent discussion in \cite{Nambu:1960p1} seems to only discuss the case where the current pattern decays with a power law as in the superfluid, rather than the superconducting case where perfect screening leads to an exponential profile as computed here.}
 (Note that the short-distance behavior of the wavepacket is qualitatively different depending on whether the superconductor is Type I (or weakly Type II) in which case $\lambda_L$ completely characterizes the current distribution, or strongly Type II, with $\xi_0 \ll \lambda\ll \lambda_L$, in which case the bogolon resembles that in a neutral superfluid for $r \ll \lambda_L$.)

 {\it Paired QH States.--} Our final example is the case of a bogolon in a paired QH state of composite fermions (CFs). Here, we start with fermions moving in a {\it static} uniform background field $\mathbf{A}$ (where $\nabla\times \mathbf{A} =  \mathbf{B}$), and perform a `flux attachment' by means of a statistical gauge field $\mathbf{a}$ whose dynamics are governed by a Chern-Simons (CS) term, $\mathcal{L}_\text{CS} = \frac{1}{4\Phi_0} \epsilon^{\mu\nu\rho} a_\mu \partial_\nu a_\rho$, with $\Phi_0$ the quantum of flux. Qualitatively, the role of the CS gauge field is to attach two quanta of magnetic flux to each  electron to convert it into a CF, which sees zero net flux at half-filling, i.e. we have $\mathbf{B} = 2\Phi_0\bar\rho$. In this case, we replace $\partial_\mu \rightarrow D_\mu = \partial_\mu - i(a+A)_\mu$, and change the currents and densities accordingly. Although more properly we should consider the example of spinless fermions and $p$-wave pairing, the distinction is unimportant as we are primarily interested in the interplay of the CS electrodynamics and charge conservation, neither of which depends essentially on the pairing symmetry. The equations of motion now follow as a result of $S +S_\text{CS}$: the `matter' equations are similar to the previous example,
\be\label{eq:eom_CS}
\rho &=& \rho^\text{qp} - \chi_0 [\partial_t \theta-2(a_0+A_0)], \\
 \mathbf{j}& = &\mathbf{j}^{ \text{qp}} +\frac{n_s}{2m} [\nabla\theta-2(\mathbf{a}+\mathbf{A})], \ \ \ \ \
\partial_t \rho^\text{qp} =  -\nabla\cdot\mathbf{j}^{ \text{qp}}  +\mathcal{B}_p\nonumber\\
\chi_0&\partial_t& [\partial_t \theta - 2(a_0+A_0)]  =\frac{n_s}{2m} \nabla\cdot[\nabla\theta -2(\mathbf{a} +\mathbf{A})] +\mathcal{B}_p \nonumber,\ee
but the Maxwell equations are replaced by the CS equations, which are pure constraints:
\be
b &\equiv& \nabla\times \mathbf{a} =  - 2\Phi_0 \left\{\rho^\text{qp} -\chi_0[\partial_t\theta - 2(a_0+A_0)]\right\} \\
 \mathbf{e} &\equiv& -\partial_t \mathbf{a} -\nabla a_0  = 2\Phi_0 \zhat\times \left\{\mathbf{j}^\text{qp} + \frac{n_s}{2m}[\nabla\theta-2(\mathbf{a}+\mathbf{A})]\right\}.\nonumber \ee

Note that now, $\mathbf{A}$ is not a dynamical field, but rather represents the background magnetic field, $  \frac{\nabla\times \mathbf{A}}{2\Phi_0}=\bar{\rho}$, where $\bar\rho$ is the mean density, $\langle \rho_F \rangle_0$, in the groundstate, and, for the present, we will set the  external potential $A_0=0$ \footnote{$A_0\neq0$ corresponds to a trapping potential for the quasiparticle-- a case which is more involved, and does not offer significantly more insight.}.
Proceeding to the wave packet state, and
specializing to unitary gauge and to static configurations as in the previous examples
we
 use the identity $\nabla^2\mathbf{V} = \zhat\times\nabla[\nabla\times \mathbf{V}] + \nabla[\nabla\cdot\mathbf{V}]$ (valid in $d=2$) to write
\be
\left[-\nabla^2 + \lambda_\text{\tiny CS}^{-2}\right](\mathbf{a}+ \mathbf{A}) = 8\chi_0 \Phi_0^2 \left\langle \mathbf{j}^\text{qp} - {\lambda_\text{\tiny CS}^{2}} \nabla(\nabla\cdot\mathbf{j}^\text{qp})\right\rangle \label{eq:CSJ}\nonumber %\\
\ee
where $\lambda_\text{\tiny CS}^{-2} =   \frac{8\chi_0\Phi_0^2  n_s}{m}$. The solution for the total field $\mathbf{a}+\mathbf{A}$ is similar to (\ref{eq:Maxkspacecurrent});
observe that the flux is exponentially screened over a distance $\lambda_\text{\tiny CS}$. Using $\chi_0 \sim \frac{m}{2\pi}$ as appropriate to a 2D Fermi surface, and the QH relation $\bar{\rho} = 1/4\pi \ell_B^2$ for filling factor $\nu=1/2$ where $\ell_B$ is the magnetic length, we find $\lambda_\text{\tiny CS} \sim\frac{1}{2} {\ell_B} \left({\bar\rho}/{n_s}\right)^{1/2}$, 
so for a Galilean invariant system at $T=0$ where $n_s =\bar\rho$, we find $\lambda_\text{\tiny CS} \sim \frac{1}{2}{\ell_B}$. Thus, the characteristic size of a bogolon wave packet is of order the magnetic length. A striking difference from the normal superconductor is that the second CS equation forces the existence of an electric field, which leads to a deviation of charge density from the background. The simplest estimate is $\delta\rho \sim \frac{1}{\ell_B^2} \frac{(\hat{\bk}_0 \times \br)
\cdot \zhat}{r}e^{-2r/\ell_B}$; while this is not the exact form, the important point is that there is {\it necessarily} a dipolar charge distribution oriented perpendicular to $\bk_0$, with separation $\sim \ell_B$, accompanying the screened dipolar current pattern. Upon inclusion of the long-range Coulomb interaction (ignored so far) \cite{SupplementaryMaterial} both current and charge densities acquire power-law tails similar to those in \cite{Sondhi:1992p1220}.

In the QH case, the bogolon has a natural interpretation as the descendant of the CF in the paired phase. Several authors, including one of us  \cite{Read:1998p1, Pasquier:1998p1, Lee:1998p1,Murthy:2003p1}, have observed that upon projection to the lowest Landau level  the CF in the compressible phase goes from being a charged particle to a neutral particle with a dipole moment proportional to its speed and perpendicular to its direction of propagation. The argument for a dipolar charge distribution for the bogolon presented here -- the application of CS electrodynamics to a paired superfluid -- is rather different from projection to a reduced Hilbert space,  and the connection between the two cases is an intriguing question  that we hope to address in the future. We note that a recent microscopic study \cite{Sreejith:2011p1}
reports an excitonic construction of the quantum Hall bogolon in the Pfaffian state
that is also consistent with an associated dipolar charge distribution.

 {\it Concluding Remarks.---}
In this paper, we have given a consistent microscopic description of bogolon wave packets in three broad classes of paired fermion states: superfluids, superconductors, and paired composite Fermi liquids with CS electrodynamics. In all cases, the quasiparticle is associated with a decidedly nontrivial current flow pattern carried in part by the condensate, and manifestly obeys global and/or local conservation laws as appropriate. Although for pedagogical simplicity we focused on the case of stationary wave packets, this restriction is merely a matter of convenience:  suitably boosted current configurations are associated with bogolons in motion.

Our results are valid in the limit $\overline{\rho^\text{qp}}\xi^{2}\ll 1$ when the concentration of quasiparticle wavepackets is small  , or in other words when the  distance between quasiparticles is much larger than their size. In the opposite limit of { high} quasiparticle concentration where quasiparticles overlap, the system can be studied using the kinetic equation approach \cite{RussianReview}. In this formalism, the Boltzmann equation  for  the distribution function of quasiparticles $n_{{\bf k}}$ is supplemented by equations of motion for the electrodynamic fields and continuity equations expressing charge conservation. The current and charge densities take the form
\be
{\bf j}&=&e\rho{\bf v}_\text{s}+e\int d^d {k}\, \frac{\bf k}{m}n_{{\bf k}}; \nonumber\\ \rho&=&e\int d ^d{k}\, [u^{2}_{{\bf k}}n_{{\bf k}}+v^{2}_{{\bf k}}(1-n_{-{\bf k}})],
\ee
where ${\bf v}_\text{s} = \frac{1}{2}\left(\nabla\theta-2\mathbf{A}\right)$ is the superfluid velocity. The situation is similar to the microscopic scenario discussed here: in order to describe the distribution of ${\bf v}_\text{s}$, an additional variable included in the kinetic theory 
 (compared to the case of the normal metal), charge conservation must be treated as an independent equation, rather than following directly from the equations of motion. 

\acknowledgements{We thank D.-H. Lee for useful discussions. BZS, SAP and SLS acknowledge the hospitality of the Stanford Institute for Theoretical Physics, where this work was initiated. This research was supported in part by the Simons Foundation (SAP) as well as NSF grants DMR-1006608 and PHY-1005429 (SAP, SLS), DMR-0901903 (RS) and DMR-0704151 (BZS), and DOE grant AC02-76SF00515 (SAK).
}

{\it Note.---} This Letter was published with the modified title {\it Microscopic Model of Quasiparticle Wave Packets in Superfluids, Superconductors, and Paired Hall States}.
\bibliography{QPCurrents-bib}
\newpage
\begin{widetext}
\begin{appendix}
\section*{Supplementary Material}
\subsection{Bogoliubov-de Gennes Equations and the BCS Ground State}
We begin with the reduced BCS Hamiltonian,
\be
H_{\text{BCS}} =  \sum_{\substack{\bk \\ s = \uparrow,\downarrow}} \xi_{\bk} c^\dagger_{\bk s} c_{\bk s} + \sum_{\bk} \Delta_\bk c^\dagger_{\bk\uparrow} c^\dagger_{-\bk\downarrow}+ \Delta_\bk^*c_{-\bk\downarrow}c_{\bk\uparrow} 
\ee

where $\xi_{\bk} = \frac{\bk^2}{2m} -\mu$, and $\Delta_0 $ is the pair potential.  As is standard, we define Bogoliubov operators via
\be
\gamma_{\bk,\uparrow} = u_\bk c_{\bk,\uparrow} - v_{\bk} c^\dagger_{-\bk,\downarrow},& &
\gamma_{\bk,\downarrow} = v_\bk c^\dagger_{\bk,\uparrow} + u_{\bk} c_{-\bk,\downarrow}  
\ee
Note that for the quasiparticle operators we have assigned spin by noting that removing a spin up and adding a spin down both lead to a state with a net spin of $-1/2$.
As usual, we determine $u_\bk, v_\bk$ by requiring the Hamiltonian to be diagonal in the quasiparticle operators, which gives
\be
|u_\bk|^2 = \frac{1}{2} \left(1 + \frac{\xi_\bk}{E_\bk}\right),  & &
|v_\bk|^2 = \frac{1}{2} \left(1 - \frac{\xi_\bk}{E_\bk}\right) \nonumber\\
\ee
where $E_\bk = \sqrt{\xi_\bk^2 + |\Delta_{\bk}|^2}$, and
\be
H = \sum_{\substack{\bk\\ s =\uparrow,\downarrow}} E_{\bk} \gamma^\dagger_{\bk s} \gamma_{\bk s}
\ee

Since $|u_\bk|^2 + |v_\bk|^2 = 1$, we may invert the Bogoliubov transformations as
\be
c_{\bk,\uparrow} = u_\bk^* \gamma_{\bk,\uparrow} +v_\bk \gamma^\dagger_{\bk,\downarrow}, & &c_{-\bk,\downarrow} = -v_\bk \gamma^\dagger_{\bk, \uparrow} + u_\bk^* \gamma_{\bk,\downarrow} 
\ee

The eigenvalues of the BdG hamiltonian come in pairs with positive and negative energies. The $T=0$ BCS ground state is obtained by filling all the negative-energy quasiparticle levels
\be
\ket{\Omega}  = \prod_{\bk \ge 0} \left(u_\bk + v_\bk c^\dagger_{\bk\uparrow}c^\dagger_{-\bk\downarrow} \right)\ket{0}
\ee
and is annihilated by all the Bogoliubov operators, $\gamma_{\bk s} \ket{\Omega} =0$ for all $\bk$.

\subsection{The Wavepacket State}
The quasiparticle dispersion has a minimum at $k = k_F$. An excited state with one quasiparticle with momentum $\bk$ and energy $E_\bk$ and spin $s$ is given by
\be
\ket{\bk s} = \gamma^\dagger_{\bk s} \ket{\Omega}
\ee

A low-energy wavepacket solution with average momentum $\bk_0$ and spatial extent $\sim \lambda$ can also be constructed, as

\be
\ket{\Psi^\lambda_{\bk_0,s}} = \left(\frac{\lambda}{\sqrt{\pi}}\right)^{\frac{d}{2}}\int{d^dk}\, e^{-\frac12\lambda^2(\bk - \bk_0)^2} \ket{\bk s}  =  \left(\frac{\lambda}{\sqrt{\pi}}\right)^{\frac{d}{2}}\int{d^dk}\,  e^{-\frac12\lambda^2(\bk - \bk_0)^2} \gamma^\dagger_{\bk s} \ket{\Omega}
\ee

with $|\bk_0| \sim k_F$, so that the energy is small. For $\left||\bk_0|-k_F\right| \ll \Delta_0$, the energy is approximately $E(\bk_0) \approx \Delta_0 + c_0\frac{ v_F^2  (|\bk_0|-k_F)^2}{\Delta_0}$ where $c_0$ is a number of order $1$.

\subsection{Current and Density in the Wavepacket State}
The current operator is given by $\mathbf{j}_\bq  = \sum_{\bk,s} \frac{\bk}{m} c^\dagger_{\bk +\frac{\bq}{2} s} c_{\bk -\frac{\bq}{2} s}$, and

\be
\bra{\Psi^\lambda_{\bk_0,s}}\mathbf{j}_\bq\ket{\Psi^\lambda_{\bk_0,s}} &=& \left(\frac\lambda{\sqrt{\pi}}\right)^d \int d^d k' \int d^d k'' e^{-\frac{1}{2}\lambda^2\left[ (\bk' -\bk_0)^2 +(\bk'' -\bk_0)^2\right]} \bra{\bk' s} \mathbf{j}_\bq \ket{\bk'' s}
\ee
We may now reduce the current matrix element in the excited state to a ground state average, using the fact that the Bogoliubov operators annihilate the BCS ground state $\ket{\Omega}$:
\be
\bra{\bk' s} \mathbf{j}_\bq \ket{\bk'' s} &=& \bra{\Omega} \gamma_{\bk's} \mathbf{j}_\bq \gamma^{\dagger}_{\bk''s}\ket{\Omega}\nonumber\\
&=& \bra{\Omega} \gamma_{\bk's}\left(\left[\mathbf{j}_\bq,  \gamma^{\dagger}_{\bk''s}\right] +  \gamma^{\dagger}_{\bk''s} \mathbf{j}_\bq\right)\ket{\Omega}\nonumber\\
&=& \bra{\Omega} \left\{ \gamma_{\bk's},\left[\mathbf{j}_\bq,  \gamma^{\dagger}_{\bk''s}\right]\right\} \ket{\Omega} + \bra{\Omega} \mathbf{j}_\bq \ket{\Omega} \delta(\bk' -\bk'')
\ee

The first term may be simplified as
\be
\left\{ \gamma_{\bk'\uparrow},\left[\mathbf{j}_\bq,  \gamma^{\dagger}_{\bk''\uparrow}\right]\right\} &=& \sum_{\bk,s} \frac{\bk}{m} \left\{u_{\bk'} c_{\bk'\uparrow} - v_{\bk'}c^\dagger_{-\bk',\downarrow}\,\,,\,\, \left[ c^\dagger_{\bk +\frac{\bq}{2} s} c_{\bk -\frac{\bq}{2} s}\,\,,\,\, u_{\bk''}^*c^\dagger_{\bk'',\uparrow} - v_{\bk'}^* c_{-\bk'',\downarrow}\right]\right\} \nonumber\\
&=& \sum_{\bk} \frac{\bk}{m} \left\{u_{\bk'} c_{\bk'\uparrow} - v_{\bk'}c^\dagger_{-\bk',\downarrow}\,\,,\,\,u^*_{\bk''} c^\dagger_{\bk+\frac{\bq}{2},\uparrow} \delta\left(\bk'' -\bk +\frac{\bq}{2}\right) + v^*_{\bk''} c_{\bk-\frac{\bq}{2},\downarrow} \delta\left(-\bk'' -\bk -\frac{\bq}{2}\right)\right\} \nonumber\\
&=& \sum_{\bk} \frac{\bk}{m} \left[u_{\bk'} u^*_{\bk''} \delta\left(\bk' -\bk -\frac{\bq}{2}\right)\delta\left(\bk'' -\bk +\frac{\bq}{2}\right) - v_{\bk'} v^*_{\bk''} \delta\left(-\bk' -\bk +\frac{\bq}{2}\right)\delta\left(-\bk'' -\bk -\frac{\bq}{2}\right)\right]\nonumber\\
&=& \frac{\bk'+\bk''}{2m} \left( u_{\bk'}u^*_{\bk''} + v_{\bk'}v^*_{\bk''} \right) \delta(\bk'-\bk'' -\bq)
\ee
and, similarly
\be
\left\{ \gamma_{\bk'\downarrow},\left[\mathbf{j}_\bq,  \gamma^{\dagger}_{\bk''\downarrow}\right]\right\}&=& - \frac{\bk'+\bk''}{2m} \left( u_{\bk'}u^*_{\bk''} + v_{\bk'}v^*_{\bk''} \right) \delta(\bk'-\bk'' +\bq)
\ee
while the second term can be made to vanish by choosing a real wavefunction for the ground state.
 
For concreteness, let us add a spin up quasiparticle. Then, the current in the wavepacket is
\be
\bra{\Psi^\lambda_{\bk_0,\uparrow}}\mathbf{j}_\bq\ket{\Psi^\lambda_{\bk_0,\uparrow}}  &=& \left(\frac\lambda{\sqrt{\pi}}\right)^d \int d^d k\, e^{-\lambda^2(\bk -\bk_0)^2 + \frac{\lambda^2 \bq^2}{4} } \frac{\bk}{m}\left( u_{\bk +\frac{\bq}{2}}u^*_{\bk - \frac{\bq}{2}} + v_{\bk +\frac{\bq}{2}}v^*_{\bk - \frac{\bq}{2}}   \right)
\ee

For the  $s$-wave case, we may take  (in the first iteration, i.e. at the mean-field level before enforcing self-consistency) the $u$s and $v$s real, in which case the purely $\bq=0$ contributions to ${\mathcal{B}_p}$ and $\mathbf{j}$ both vanish.
For specificity, we work in two dimensions and  choose a stationary wavepacket with $\bk_0 =  k_F \hat{\mathbf{x}}$, which corresponds to a quasiparticle at rest\footnote{Note that in the main body of the paper we rewrite this in coordinate independent fashion, but for the supplementary material it is convenient to work with an explicit choice for $\bk_0$.}. If we define $\bk = \bk_0 + \tilde{\bk}$, we have
\be
|u_{\bk\pm \frac{\bq}{2}}| \approx \frac{1}{\sqrt{2}}\left\{ 1 + \frac{v_F\left( \tilde{k}_x \pm \frac{q_x}{2}\right)}{2\Delta_0}\right\} + \mathcal{O}(\tilde{k}^2), & &
|v_{\bk\pm \frac{\bq}{2}}| \approx \frac{1}{\sqrt{2}}\left\{ 1 - \frac{v_F\left( \tilde{k}_x \pm \frac{q_x}{2}\right)}{2\Delta_0}\right\} + \mathcal{O}(\tilde{k}^2) \label{eq:uvdef}
\ee 
which, when substituted into the expression of the current (with $\bk$ shifted as above) yields
\be
\langle\mathbf{j}_\bq\rangle_\Psi &=&e^{- \frac{\lambda^2 \bq^2}{4} }\left(\frac\lambda{\sqrt{\pi}}\right)^2 \int d^2 \tilde{k}\, e^{-\lambda^2\tilde{k}^2}\frac{1}{m} \left(k_F\hat{\mathbf{x}} +\tilde\bk\right) \nonumber\\ & & \times\left[\frac{1}{2} \left\{ 1 + \frac{v_F\left( \tilde{k}_x + \frac{q_x}{2}\right)}{2\Delta_0}\right\} \left\{ 1 + \frac{v_F\left( \tilde{k}_x - \frac{q_x}{2}\right)}{2\Delta_0}\right\} +\frac{1}{2} \left\{ 1 - \frac{v_F\left( \tilde{k}_x + \frac{q_x}{2}\right)}{2\Delta_0}\right\} \left\{ 1 - \frac{v_F\left( \tilde{k}_x - \frac{q_x}{2}\right)}{2\Delta_0}\right\} \right] \nonumber\\
&\approx&  v_F e^{- \frac{\lambda^2 \bq^2}{4} }\hat{\mathbf{x}}  \label{eq:wpcurrent}
\ee
For the density in the wave packet state, a similar calculation yields
\be
\bra{\Psi^\lambda_{\bk_0,\uparrow}}\rho_\bq\ket{\Psi^\lambda_{\bk_0,\uparrow}}  &=&  \bar{\rho}\delta_{\bq,0}  + \left(\frac\lambda{\sqrt{\pi}}\right)^d \int d^d k\, e^{-\lambda^2(\bk -\bk_0)^2 + \frac{\lambda^2 \bq^2}{4} }\left( u_{\bk +\frac{\bq}{2}}u^*_{\bk - \frac{\bq}{2}} - v_{\bk +\frac{\bq}{2}}v^*_{\bk - \frac{\bq}{2}}   \right) \approx  \bar{\rho}\delta_{\bq,0}.
\ee

\subsection{Backflow Source Term}
Th source term to compute the backflow current is given by the expectation value of the operator
\be
{\mathcal{B}_p}(\br)&=& 2i \Delta_0  \left\{\psi^\dagger_\uparrow(\br)\psi^\dagger_\downarrow(\br) - \psi_\downarrow(\br)\psi_\uparrow(\br) \right\} \nonumber\\
&=& 2i\Delta_0 \sum_{\bk\bk'}\left\{ c^\dagger_{\bk,\uparrow} c^\dagger_{-\bk',\downarrow} e^{i (\bk - \bk')\cdot \br} - c_{-\bk',\downarrow} c_{\bk,\uparrow} e^{-i(\bk-\bk')\cdot\br}\right\} \nonumber\\
&=& 2i\Delta_0 \sum_{\bk\bk'}\left\{ \left(u_\bk \gamma^\dagger_{\bk,\uparrow} + v_\bk^* \gamma_{\bk,\downarrow} \right)\left( -v_{\bk'}^* \gamma_{\bk', \uparrow} + u_{\bk'} \gamma^\dagger_{\bk',\downarrow}\right) e^{i (\bk - \bk')\cdot \br} \right.\nonumber\\ & &\left.\,\,\,\,\,\,\,\,\,\,\,\,\,\,\,\,\,\,\,\,\,\,- \left(-v_{\bk'} \gamma^\dagger_{\bk', \uparrow} + u_{\bk'}^* \gamma_{\bk',\downarrow} \right)\left(u_\bk^* \gamma_{\bk,\uparrow} +v_\bk \gamma^\dagger_{\bk,\downarrow}\right) e^{-i(\bk-\bk')\cdot\br}\right\} \nonumber\\ &=& 2i\Delta_0 \sum_{\bk\bk'}\left\{ \left(u_\bk \gamma^\dagger_{\bk,\uparrow} + v_\bk^* \gamma_{\bk,\downarrow} \right)\left( -v_{\bk'}^* \gamma_{\bk', \uparrow} + u_{\bk'} \gamma^\dagger_{\bk',\downarrow}\right)- \left(-v_{\bk} \gamma^\dagger_{\bk, \uparrow} + u_{\bk}^* \gamma_{\bk,\downarrow} \right)\left(u_{\bk'}^* \gamma_{\bk',\uparrow} +v_{\bk'} \gamma^\dagger_{\bk',\downarrow}\right)\right\}  e^{i (\bk - \bk')\cdot \br} \nonumber\\
&=& 2i\Delta_0 \sum_{\bk\bk'} \left\{ \left(u^*_{\bk'} v_\bk - v^*_{\bk'} u_\bk \right)\gamma^\dagger_{\bk,\uparrow}\gamma_{\bk',\uparrow} + \left(v_{\bk}^* u_{\bk'} - u_{\bk}^*v_{\bk'} \right)\gamma_{\bk,\downarrow}\gamma^\dagger_{\bk',\downarrow} \right.\nonumber\\ & &\left.\,\,\,\,\,\,\,\,\,\,\,\,\,\,\,\,+ \left(u_\bk u_{\bk'} +v_\bk v_{\bk'}\right)\gamma^\dagger_{\bk,\uparrow}\gamma^\dagger_{\bk',\downarrow} +\left( -u_\bk^*u_{\bk'}^* - v_\bk^* v_{\bk'}^*\right) \gamma_{\bk,\downarrow}\gamma_{\bk',\uparrow}\right\}e^{i(\bk-\bk')\cdot \br}
\ee
where in going from the third to the fourth line we have interchanged $\bk, \bk'$ in the second term.

Proceeding as before, we compute expectation values of the source term in the wavepacket state, and find that only the first term has a generic contribution; the second term contributes only when $\bk'=\bk''$, while the last two vanish:
\be
\langle 
{\mathcal{B}_p}(\br)\rangle_\Psi &=&\left(\frac\lambda{\sqrt{\pi}}\right)^d \int d^dk'd^dk'' \, e^{-\frac{1}{2}\lambda^2\left[ (\bk' -\bk_0)^2 +(\bk'' -\bk_0)^2\right]} \bra{\bk' s} {\mathcal{B}_p}(\br) \ket{\bk'' s} \nonumber\\
 &=&2i\Delta_0\left(\frac\lambda{\sqrt{\pi}}\right)^d \int d^dk'd^dk'' \, e^{-\frac{1}{2}\lambda^2\left[ (\bk' -\bk_0)^2 +(\bk'' -\bk_0)^2\right]} \left[\left( u^*_{\bk''} v_{\bk'} - v^*_{\bk''}u_{\bk'}\right)\right. \nonumber\\& &\left.\,\,\,\,\,\,\,\,\,\,\,\,\,\,\,\,\,\,\,\,\,\,\,\,\,\,\,\,\,\,\,\,\,\,\,\,\,\,\,\,\,\,\,\,\,\,\,\,\,\,\,\,\,\,\,\,\,\,\,\,\,\,\,\,\,\,\,\,\,\,\,\,\,\,\,\,\,\,\,\,\,\,\,\,\,\,\,\,\,\,\,\,\,\,\,\,\,\,\,\,\,\,\,\,\,+ \delta(\bk'-\bk'')\left(v_{\bk'}^*u_{\bk''} - u^*_{\bk'}v_{\bk''}\right)\right]e^{i(\bk'-\bk'')\cdot\br}
\ee

We may rewrite this in momentum space, $\langle{\mathcal{B}_p}(\br)\rangle_\Psi = \int \frac{d^d q}{(2\pi)^d}\langle{\mathcal{B}_p}_\bq\rangle_\Psi e^{-i\bq\cdot\br}$, whence
\be
\langle{\mathcal{B}_p}_\bq\rangle_\Psi &=&2i\Delta_0\left(\frac\lambda{\sqrt{\pi}}\right)^d \int d^dk\, e^{-\lambda^2(\bk -\bk_0)^2 - \frac{\lambda^2 \bq^2}{4} }  \left[\left( u^*_{\bk+\frac{\bq}{2}} v_{\bk-\frac{\bq}{2}} - v^*_{\bk+\frac{\bq}{2}}u_{\bk-\frac{\bq}{2}}\right) + \delta(\bq)\left(v_{\bk-\frac{\bq}{2}}^*u_{\bk+\frac{\bq}{2}} - u^*_{\bk-\frac{\bq}{2}}v_{\bk+\frac{\bq}{2}}\right)\right]\nonumber\\
\ee

Using the expressions (\ref{eq:uvdef}) for $u$ and $v$ and for simplicity \footnote{The case for $d=3$ is straightforward due to the azimuthal symmetry, but is more notationally cumbersome.}  working in $d=2$ we have, explicitly
\be
\langle{\mathcal{B}_p}_\bq\rangle_\Psi &=&2i\Delta_0 e^{- \frac{\lambda^2 \bq^2}{4} }\left(\frac\lambda{\sqrt{\pi}}\right)^2 \int d^2\tilde{k}\, e^{-\lambda^2\tilde{k}^2} \nonumber\\ & & \times\left[\frac{1}{2} \left\{ 1 + \frac{v_F\left( \tilde{k}_x + \frac{q_x}{2}\right)}{2\Delta_0}\right\} \left\{ 1 - \frac{v_F\left( \tilde{k}_x - \frac{q_x}{2}\right)}{2\Delta_0}\right\} -\frac{1}{2} \left\{ 1 - \frac{v_F\left( \tilde{k}_x + \frac{q_x}{2}\right)}{2\Delta_0}\right\} \left\{ 1 +\frac{v_F\left( \tilde{k}_x - \frac{q_x}{2}\right)}{2\Delta_0}\right\} \right]\nonumber\\
&\approx&  i q_x v_F e^{- \frac{\lambda^2 \bq^2}{4} } \label{eq:bfsrc}
\ee

\subsection{Gap Function and Phase Texturing in the Wavepacket State}
We now show that the expectation value of the gap operator  acquires a phase winding in position space when it is computed in the wavepacket state built from the uniform BdG solutions above. We have
\be
\Delta^*(\br) &=& \psi^\dagger_\uparrow(\br) \psi^\dagger_\downarrow(\br) \nonumber\\
&=&\sum_{\bk\bk'}c^\dagger_{\bk,\uparrow} c^\dagger_{-\bk',\downarrow} e^{i (\bk - \bk')\cdot \br} \nonumber\\
&=& \sum_{\bk\bk'} \left(u_\bk \gamma^\dagger_{\bk,\uparrow} + v_\bk^* \gamma_{\bk,\downarrow} \right)\left( -v_{\bk'}^* \gamma_{\bk', \uparrow} + u_{\bk'} \gamma^\dagger_{\bk',\downarrow}\right) e^{i (\bk - \bk')\cdot \br}\nonumber\\
&=& \sum_{\bk\bk'} \left\{-v_{\bk'}^*u_\bk \gamma^\dagger_{\bk,\uparrow}\gamma_{\bk', \uparrow} + v_\bk^*u_{\bk'} \gamma_{\bk,\downarrow} \gamma^\dagger_{\bk',\downarrow}
  + u_\bk u_{\bk'} \gamma^\dagger_{\bk,\uparrow}  \gamma^\dagger_{\bk',\downarrow} - v_\bk^*v_{\bk'}^* \gamma_{\bk,\downarrow}  \gamma_{\bk', \uparrow}
  \right\}e^{i(\bk -\bk')\cdot \br}
\ee
In the wavepacket state, we proceed as in the previous section and find as before that the latter two terms vanish, the second is a purely $\bq=0$ contribution, while the first term is `generic'. Going to momentum space as before we have,
\be
\langle \Delta^*_\bq\rangle_\Psi &=& \left(\frac\lambda{\sqrt{\pi}}\right)^d \int d^dk\, e^{-\lambda^2(\bk -\bk_0)^2 - \frac{\lambda^2 \bq^2}{4} }  \left[ - v^*_{\bk+\frac{\bq}{2}}u_{\bk-\frac{\bq}{2}} + \delta(\bq)v_{\bk-\frac{\bq}{2}}^*u_{\bk+\frac{\bq}{2}}\right]
\ee
Exactly at $\bq=0$,  the two terms cancel so that the expectation value vanishes. For $\bq \neq 0$ on the other hand we may drop the second term and write ($d=2$) as before
\be
\langle \Delta^*_\bq\rangle_\Psi &=&-\left(\frac\lambda{\sqrt{\pi}}\right)^2 \int d^2k\, e^{-\lambda^2(\bk -\bk_0)^2 - \frac{\lambda^2 \bq^2}{4} }  v^*_{\bk+\frac{\bq}{2}}u_{\bk-\frac{\bq}{2}} \nonumber\\
&=& -\frac{1}{2}{e^{ - \frac{\lambda^2 \bq^2}{4} }  }\left(\frac\lambda{\sqrt{\pi}}\right)^2 \int d^2\tilde{k}\, e^{-\lambda^2 \tilde{k}^2} \left\{ 1 - \frac{v_F\left( \tilde{k}_x + \frac{q_x}{2}\right)}{2\Delta_0}\right\} \left\{ 1 +\frac{v_F\left( \tilde{k}_x - \frac{q_x}{2}\right)}{2\Delta_0}\right\} \nonumber\\
&=& -\frac{1}{2}{e^{ - \frac{\lambda^2 \bq^2}{4} }  }\left(\frac\lambda{\sqrt{\pi}}\right) \int d\tilde{k}_x\, e^{-\lambda^2 \tilde{k}_x^2} \left\{ \left(1- \frac{v_F q_x}{4\Delta_0}\right)^2 - \frac{v_F^2}{4\Delta_0^2} \tilde{k}_x^2 \right\} \nonumber\\
&=&  -\frac{1}{2}  \left\{ \left(1- \frac{v_F q_x}{4\Delta_0}\right)^2 - \frac{v_F^2}{8\Delta_0^2 \lambda^2}  \right\}{e^{ - \frac{\lambda^2 \bq^2}{4} }  }
\ee
Transforming to position space, we have
\be
\langle \Delta^*(\br) \rangle_\Psi &=& \int \frac{d^2q}{(2\pi)^2} \langle \Delta^*_\bq\rangle_\Psi  e^{-i\bq\cdot\br}\nonumber\\
						&=& -\frac{1}{8\pi^2}\int {d q_y} \, e^{-i q_y y - \frac{\lambda^2 q_y^2}{4}}  \int dq_x\,  \left\{ \left(1- \frac{v_F q_x}{4\Delta_0}\right)^2 - \frac{v_F^2}{8\Delta_0^2 \lambda^2}  \right\}e^{-iq_x x - \frac{\lambda^2 q_x^2}{4} } \nonumber\\
&=& -\frac{1}{2\pi \lambda^2 }\left\{ \left(1- \frac{v_F i \partial_x}{4\Delta_0}\right)^2 - \frac{v_F^2}{8\Delta_0^2 \lambda^2}   \right\} e^{- r^2/\lambda^2}\nonumber\\
&=& -\frac{1}{2\pi \lambda^2} \left(1 + i \frac{v_F x}{2\Delta_0 \lambda^2}\right)^2e^{-r^2/\lambda^2}
\ee
From this we find the phase of the order parameter has the form (recall $\Delta_0^*  = e^{-i\theta}$) 
\be
\theta(\br) =- 2 \tan^{-1} \frac{v_F x}{2\Delta_0 \lambda^2} - \pi
\ee
which gives rise to a current in the $-\hat{\bx}$ direction, opposite to the quasiparticle current. Upon iteration of the BdG equations, the gap function textures in an attempt to reestablish charge conservation.

\subsection{Continuity-consistent solutions: quasiparticle wavepackets with backflow}

As demonstrated in the text, the total current has a dipolar form, 
\be
\mathbf{j}_{\bq}=\left\langle\mathbf{j}^\text{F}_\bq\right\rangle_\Psi + \frac{i\bq}{q^2} \langle{\mathcal{B}_p}_\bq\rangle_\Psi = v_F e^{- \frac{\lambda^2 \bq^2}{4} } \hat{\mathbf{x}}\cdot\mathcal{P}_{\mathbf{q}}
\ee

where $\mathcal{P}_{ij} = \left( \delta_{ij} - \frac{q_iq_j}{q^2}\right)$ is the transverse projector.

To determine the current flow pattern in position space, we simply Fourier transform,
\be
\mathbf{j}^{\rm T}(\br) = \int \frac{d^2 q}{(2\pi)^2}\, \mathbf{j}^{\rm T}_\bq e^{-i \bq\cdot\br} &=&   \frac{v_F}{(2\pi)^2} \int _{0}^{2\pi}d\theta \int_0^\infty d q\,  \frac{1}{q}  e^{-\frac{\lambda^2 q^2}{4}} e^{-i \bq\cdot\br}  \left[\hat{\mathbf{x}}{ q_y^2}  - \hat{\mathbf{y}} {q_x q_y } \right] \nonumber\\
&=&  - \frac{v_F}{(2\pi)^2}  \left[\hat{\mathbf{x}}\left(\frac{\partial}{\partial y}\right)^2 - \hat{\mathbf{y}} \frac{\partial^2}{\partial x \partial y} \right] \int_0^\infty d q\,  \frac{1}{q}  e^{-\frac{\lambda^2 q^2}{4}} \int _{0}^{2\pi}d\theta  e^{-i qr \cos\theta}  \nonumber\\
&=&  \frac{v_F}{(2\pi)^2} \left[\hat{\mathbf{y}} \partial_x -  \hat{\mathbf{x}} \partial_y\right] \frac{\partial}{\partial y}  \left[\int_0^\infty d q\,  \frac{J_0(qr)}{q}  e^{-\frac{\lambda^2 q^2}{4}} \right] \nonumber\\
&=& \frac{v_F}{2\pi}\hat{\mathbf{z}}\times \nabla \left[ - \frac{y}{r} \int_{0}^\infty dq\, J_1(qr)  e^{-\frac{\lambda^2 q^2}{4}} \right] \nonumber\\
&=& -  \frac{v_F}{2\pi} \hat{\mathbf{z}}\times \nabla \left[ \frac{y}{r^2} \left(1 - e^{-r^2/\lambda^2}\right)\right] \nonumber\\
&=&  \left[\hat{\mathbf{x}} \partial_y  - \hat{\mathbf{y}} \partial_x \right] \varphi_\lambda(\br)  
\ee
where we defined $\varphi_\lambda(\br) = \frac{v_F}{2\pi} \frac{y}{r^2}\left(1 - e^{-r^2/\lambda^2}\right)$. The flow pattern is solenoidal, and falls off as $1/r^2$ far away from the center of the wavepacket.

\subsection{Long-Range Current Flow Pattern from the Hall Conductance}

We have observed (see main text) that in the QH case the CS electrodynamics gives us a density of dipolar form, $\delta\rho(\br) = \frac{1}{\ell_B^2} \frac{y}{r} e^{-2r/\ell_B}$. Using  the multipole expansion for the Coulomb interaction $v(\br) =e^2/r$, we have for the electric potential due to the quasiparticle wavepacket
\be
V(r,\theta) = \int d\br' \delta\rho(\br') \frac{e}{|\br -\br'|} &=& \sum_{l=0}^{\infty}\frac{1}{r^{l+1}} \int d\br' (r')^l (\delta\rho(\br')) P_l(\cos(\theta-\theta'))
\ee
Clearly by symmetry the monopole contribution (the $l=0$ term) vanishes. The dipole ($l=1$) term is nonzero, and thus the leading term in $V$ is (using coordinates where angles are measured with respect to the $x$-axis)
\be
V(r,\theta) &\approx& \frac{e}{\ell_B^2 r^2}\int_0^\infty r'^2 dr' e^{-2r'/\ell_B}  \int _0^{2\pi} d\theta' \sin(\theta') (\cos(\theta-\theta'))\nonumber\\
 &=& \frac{e\ell_B}{4r^2} \int _0^{2\pi} d\theta' \cos(\theta') (\cos\theta\cos\theta' +\sin\theta\sin\theta') \nonumber\\
 &=&  \frac{\pi}{4} e\ell_B \frac{\sin\theta}{r^2}
\ee
This yields an electric  field of the form (valid asymptotically far  from the wave packet center)
\be
\mathbf{E} = -\nabla V(r,\theta)\propto \frac{ e\ell_B}{ r^3} \left(2\sin\theta\hat{\br} - \cos\theta\hat{\boldsymbol{\theta}}\right) \ee
corresponding to a dipole moment of $\pi e \ell_B/4$ (note that in standard references on electrostatics, this result is often quoted  with angles referred to the dipole axis, which is perpendicular to the choice  here.)
From this, using the fact that we are in a quantized Hall state with $\sigma_{xy}= \frac{e^2}{2h}$ we find a current flow pattern at long distances that takes the form
\be
\mathbf{j} = \frac{e^2}{2h} \zhat\times \mathbf{E}  = \frac{e^3\ell_B}{8h} \frac{\left(2\sin\theta\hat{\boldsymbol{\theta}} + \cos\theta\hat{\br}\right)}{r^3} \ee
Thus, there is an asymptotic power law tail in the current density due to the long-range nature of the Coulomb interaction (see also Ref. \cite{Sondhi:1992p1220} of the main text.)  The important points to note are the power-law form of the current and its dipolar nature, rather than the numerical prefactor, which is not precise since we use only the most naive estimate of the bogolon charge density. Further refinements will produce smaller, power law corrections to the charge density
profile as well.
\end{appendix}
\end{widetext}
\end{document}